\documentclass[preprint,groupedaddress,amsmath,amssymb,superscriptaddress]{revtex4-1}
\usepackage{graphicx}
\usepackage{dcolumn}
\usepackage{bm}

\begin{document}
\title{Anomalous Itinerant-Electron Metamagnetic Transition in the Layered $\bm{\mathrm{Sr}_{1-x}\mathrm{Ca}_x\mathrm{Co}_2\mathrm{P}_2}$ System}

\author{Masaki Imai}
\email[]{m.imai@kuchem.kyoto-u.ac.jp}
\affiliation{Department of Chemistry, Graduate School of Science, Kyoto University, Kyoto 606-8502, Japan}
\author{Chishiro Michioka}
\affiliation{Department of Chemistry, Graduate School of Science, Kyoto University, Kyoto 606-8502, Japan}
\author{Hiroto Ohta}
\affiliation{Department of Applied Physics, Tokyo University of A \& T, Koganei, Tokyo 184-8588, Japan}
\author{Akira Matsuo}
\affiliation{Institute for Solid State Physics, The University of Tokyo, Kashiwanoha, Kashiwa, Chiba 277-8581, Japan}
\author{Koichi Kindo}
\affiliation{Institute for Solid State Physics, The University of Tokyo, Kashiwanoha, Kashiwa, Chiba 277-8581, Japan}
\author{Hiroaki Ueda}
\affiliation{Department of Chemistry, Graduate School of Science, Kyoto University, Kyoto 606-8502, Japan}
\author{Kazuyoshi Yoshimura}
\email[]{kyhv@kuchem.kyoto-u.ac.jp}
\affiliation{Department of Chemistry, Graduate School of Science, Kyoto University, Kyoto 606-8502, Japan}

\date{\today}

\begin{abstract}
We report magnetic properties of the layered itinerant system, Sr$_{1-x}$Ca$_x$Co$_2$P$_2$ in the magnetic field up to 70~T.
As for the exchange-enhanced Pauli paramagnetic metal SrCo$_2$P$_2$, the magnetization curve shows two characteristic anomalies. 
The low-field anomaly is small without obvious hysteresis, and the high-field one is a typical behavior of the itinerant-electron metamagnetic transition (IEMT). 
Such a successive transition in the magnetization curve cannot be explained by the conventional phenomenological theory for IEMT due to the Landau expansion of the free energy, but by the extended Landau expansion theory with distinguishable two energy states.
In the systematical study of Sr$_{1-x}$Ca$_x$Co$_2$P$_2$, furthermore, the metamagnetic transition field decreases and goes to zero as $x$ increases up to 0.5, indicating that the ferromagnetic quantum critical point (QCP) exists at $x \sim 0.5$.

\end{abstract}
\pacs{75.10.Lp ,75.30.Kz, 75.45.+j }
\maketitle

Up to now, more than 500 layered $AT_2X_2$-type compounds ($A$: alkaline metal, alkaline earth metal, lanthanide, $T$: transition metal, $X$: metalloid) with the ThCr$_2$Si$_2$ structure (space group: $I4/mmm$) have been discovered~\cite{CRYSTMET}, and found to show a wide variety of interesting physical properties such as heavy-fermion superconductivity in CeCu$_2$Si$_2$~\cite{CeCu2Si2_79}, high-$T_{\rm c}$ superconductivity and nematic hidden ordering in Ba$_{1-x}$K$_x$Fe$_2$As$_2$ and BaFe$_2$(As$_{1-x}$P$_x$)$_2$~\cite{Rotter08,Kasahara12} as well as itinerant-electron ferromagnetism in LaCo$_2$P$_2$~\cite{Reehuis94}.
These properties are highly related to their quasi-two-dimensional (2D) electronic structure, in which the electronic correlation with low-dimensional fluctuation effects becomes strong.

$AT_2X_2$ is formed by stacks of $A$ and $T_2X_2$ layers alternately and has two types of the interlayer staking bond depending on $A$ and $X$: one leads so called collapsed tetragonal (cT) structure and the other uncollapsed tetragonal (ucT) one, which categorized by the strength of the $X$-$X$ chemical bond between neighboring $T_2X_2$ layers~\cite{Hoffmann85}. 
In the case of ucT compounds, $T_2X_2$ layers are well isolated and their electronic structures are expected to have strong two-dimensionality, while in the case of cT compounds, a strong interlayer interaction would induce three-dimensional characters.
Therefore, the strength of the $X$-$X$ bond would be a key for controlling the dimensionality and physical properties of $AT_2X_2$.

Since it is known that the structure changes from ucT to cT around $x=0.5$ as $x$ increases~\cite{Jia09} in the Sr$_{1-x}$Ca$_x$Co$_2$P$_2$ system, this system should be one of the ideal systems to control the dimensionality of the magnetic interaction and the itinerancy systematically. 
The SrCo$_2$P$_2$ with the ucT cell is an enhanced Pauli paramagnetic metal without any magnetic orderings~\cite{Morsen88}. 
In the ucT region ($x \leq 0.5$) in which the electronic structure is expected to be quasi-2D, the Weiss temperature changes from negative value to 0 K with increasing $x$, suggesting that Sr$_{1-x}$Ca$_x$Co$_2$P$_2$ approaches to a ferromagnetic quantum critical point (QCP).
In the cT region ($x>0.5$) in which Sr$_{1-x}$Ca$_x$Co$_2$P$_2$ shows magnetic orderings~\cite{Jia09}, the intralayer ferromagnetic moments are coupled antiferromagnetically via layer by layer P-P bondings.

In this paper, we show a novel itinerant-electron metamagnetic transition (IEMT) in the Sr$_{1-x}$Ca$_x$Co$_2$P$_2$ system, interpret
their characteristic magnetization curves by means of the extended Landau expansion theory, and discuss the itinerant-electronic state in the quasi-2D transition metal pnictide system.

 Single crystalline and polycrystalline Sr$_{1-x}$Ca$_x$Co$_2$P$_2$ samples were prepared from Sr(2N), Ca(2N5), Co(3N) and P(red, 5N).
Pure single crystals of $x=0$, $1$ were obtained by a tin flux method~\cite{Reehuis94}.
High purity polycrystalline samples were synthesized through the following steps.
First, SrP, CaP and Co$_2$P were synthesized by heating stoichiometric mixtures of the elements. 
Obtained SrP, CaP and Co$_2$P powders were mixed at the ratio of $1.2x : 1.2(1-x) : 1.0$. The mixture was pelletized and sealed into an evacuated silica tube and then heated up to $1000\ {}^\circ\mathrm{C}$. 
To obtain high homogeneities, we repeated this process at least twice. 
Excess SrP and CaP were dissolved in water.

X-ray diffraction (XRD) patterns were measured using Cu K$\alpha$ radiation, and refined by the Le Bail method, using a computer program RIETAN-FP~\cite{Izumi07}.
The temperature dependent magnetizations of Sr$_{1-x}$Ca$_x$Co$_2$P$_2$ were measured by a Quantum Design MPMS-XL system in Research Center for Low Temperature and Materials Sciences, Kyoto University.
Magnetization curves beyond 70 T were measured by using an induction method with a multilayer pulsed magnet at the ultrahigh magnetic field laboratory of the Institute for Solid State Physics, the University of Tokyo. 

\begin{figure}
\includegraphics[width=8cm]{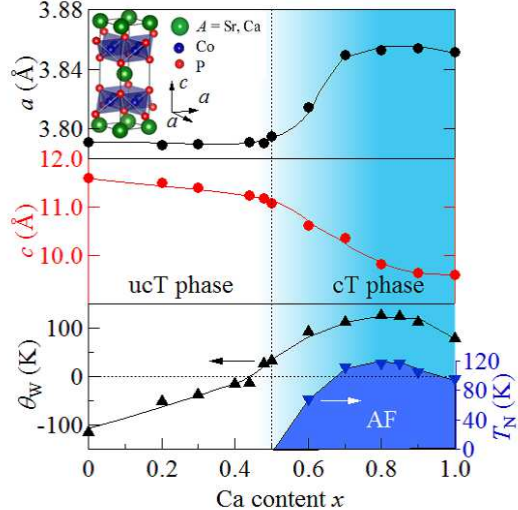}
\caption{Lattice parameters and magnetic properties of Sr$_{1-x}$Ca$_x$Co$_2$P$_2$. Lattice parameters $a$ and $c$ are shown in the upper and middle panels, respectively. Weiss and N\'eel temperatures are shown in the bottom panel. Inset: Unit cell of Sr$_{1-x}$Ca$_x$Co$_2$P$_2$. AF: antiferromagnetic state.\label{Fig:1}}
\end{figure}

Lattice parameters of Sr$_{1-x}$Ca$_x$Co$_2$P$_2$ determined from X-ray diffraction measurements are shown in Fig.~\ref{Fig:1}.
The parameter of $c$ is linearly related to the P-P distance between neighboring Co$_2$P$_2$ layers~\cite{Jia09}.
In the ucT region of $x\leq0.5$, while $a$ is almost constant, $c$ decreases monotonically with $x$. 
With decreasing the P-P distance, interlayer couplings in the quasi-2D electronic structure is enhanced. 
At $x$ = 0.5, $a$ and $c$ show a marked change according to the lattice collapse driven by forming of P-P bonds between neighboring Co$_2$P$_2$ layers.
With further increase of $x$, $c$ decreases and $a$ increases in the cT region.

\begin{figure}
\begin{center}
\includegraphics[width=8.8 cm]{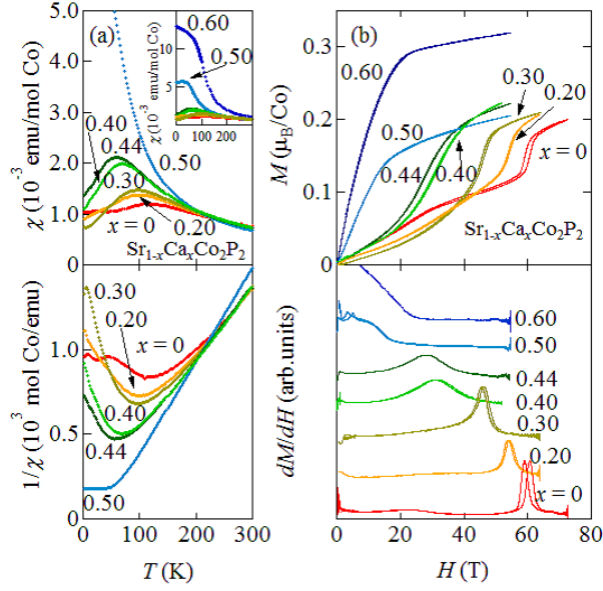}
\caption{Magnetic properties of Sr$_{1-x}$Ca$_x$Co$_2$P$_2$. (a)The magnification figure for the temperature dependence of the magnetic susceptibility and their inverses, inset: full scale. The susceptibility shows maximum behaviors in the region of $0 \leq x \leq 0.5$ without any magnetic orderings, and the maximum value of inverses as $x$ increases. (b) The magnetization $M$ and the differential magnetization $dM/dH$ measured at 4.2~K in pulsed high magnetic fields up to 70~T.\label{Fig:2}}
\end{center} 
\end{figure}

The temperature dependence of the magnetic susceptibilities $\chi(T)$ of Sr$_{1-x}$Ca$_x$Co$_2$P$_2$ is shown in Fig.~\ref{Fig:2}(a).
In all compositions, $\chi(T)$ shows a Curie-Weiss-like temperature dependence at high temperatures. 
The Weiss temperature $\theta_{\rm W}$ determined by fitting $\chi(T)$ with the Curie-Weiss formula is plotted in the lower panel in Fig.~1.
In the case of the itinerant-electron model, the origin of the Curie-Weiss-like behavior is attributed to the temperature dependence of the amplitude of local spin density, and an apparent negative Weiss temperature does not always mean the existence of antiferromagnetic interactions. 
In the case of itinerant magnetism with ferromagnetic spin fluctuations, $\theta_{\rm W}$ indicates the distance from the quantum critical point with $\theta_{\rm W} = 0$~\cite{Takahashi86, Moriya91}.
For example, nearly ferromagnetic metals such as Pd, YCo$_2$ and Sr$_{1-x}$Ca$_x$RuO$_3$ also exhibit Curie-Weiss-like temperature dependence with negative $\theta_{\rm W}$'s~\cite{Pd52,Yoshimura88,Kiyama98}.
In the Sr$_{1-x}$Ca$_x$Co$_2$P$_2$ system, $\theta_{\rm W}$ increases from negative to 0 as $x$ increases from 0 to 0.5, suggesting that ferromagnetic spin-fluctuations are enhanced and the system approaches to the QCP.

In the case of nearly ferromagnetic compounds, $\chi(T)$ often shows a maximum at a finite temperature~\cite{Yoshimura88, Goto97}.
In some compounds of Sr$_{1-x}$Ca$_x$Co$_2$P$_2$, $\chi(T)$ shows double maxima. 
In the end compound $x = 0$, the maximum of lower temperature, $T_{\mathrm{max}1}$, and that of higher temperature, $T_{\mathrm{max}2}$, are 25 and 110 K, respectively.
In our preliminary NMR study, such behaviors are found to be intrinsic.
In $x = 0.2$, $T_{\mathrm{max}1}$ and $T_{\mathrm{max}2}$ are 30 and 97.5 K, respectively. 
In $0.3 \leq x \leq 0.5$, $T_{\mathrm{max}2}$ becomes smaller as $x$ increases and $T_{\mathrm{max}1}$ disappears.
In $0.6 \leq x$, the ground state becomes an antiferromagnetic one with an antiferromagnetic interlayer coupling due to the enhanced interlayer P-P coupling.

\begin{figure}
\includegraphics[width=8.8cm]{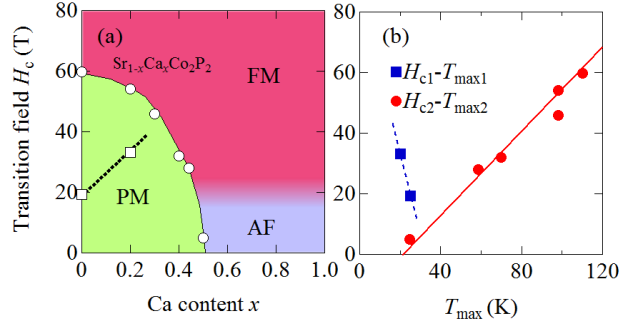}
\caption{(a) Magnetic phase diagram of Sr$_{1-x}$Ca$_x$Co$_2$P$_2$ at 4.2~K. PM: paramagnetic state; AF: antiferromagnetic state; FM: field-induced ferromagnetic state. (b) The relation between the susceptibility maximum temperatures $T_{\rm max}$ and metamagnetic transition fields $H_{\rm c}$ in the Sr$_{1-x}$Ca$_x$Co$_2$P$_2$ system. \label{Fig:3}}
\end{figure}

The magnetization $M$ and their differential $dM/dH$ curves of Sr$_{1-x}$Ca$_x$Co$_2$P$_2$ at 4.2~K are shown in Fig.~\ref{Fig:2}(b).
For $x = 0$, with increasing applied magnetic field $H$, a small anomaly of a maximum behavior in $dM/dH$ vs.~$H$ appears at $H_{\rm c1}$ = 23~T, and then an obvious peak is observed at $H_{\rm c2}$ = 59.7~T.
The latter anomaly is typical behavior of IEMT.
A hysteresis loop in $M$-$H$ curves suggests the present IEMT is a first-order transition.
For $x = 0.2$, although the anomaly at $H_{\rm c1}$ = 33~T is small, a similar behavior for the magnetization curve with $x = 0$ compound is observed.
In $0.3 \leq x \leq 0.5$, only single clear anomaly corresponding to IEMT at $H_{\rm c2}$ is observed.
The critical field $H_{\rm c2}$ becomes smaller and the anomaly becomes blunt as $x$ increases.
With increasing $x$, $H_{\rm c1}$ becomes larger and then seems to merge with $H_{\rm{c}2}$.
Within the ucT region of $0 \leq x \leq 0.5$, saturation magnetizations are about 0.2 $\mu_{\rm B}$ in all compounds. 
On the other hand, in the case of $x = 0.6$ compound which has the cT structure, the saturation magnetization is 0.3 $\mu_{\rm B}$ and larger than that in the ucT compounds.
The electronic state of P changes from isolated P$^{3-}$ to diatomic P$_2^{4-}$ with the lattice collapse~\cite{Hoffmann85}.
As a result, the hypothetical valence of Co changes and the local spin density on Co becomes larger.
From above systematical studies, we obtain the magnetic phase diagram as shown in Fig.~\ref{Fig:3}(a).

\begin{figure}
\includegraphics[width=8.8cm ]{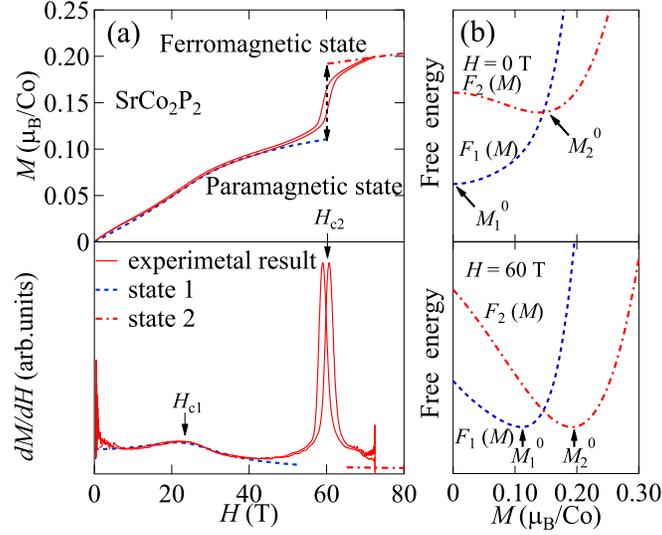}
\caption{(a) Experimental and simulated magnetization curves and their differentials of SrCo$_2$P$_2$. (b) Schematic free energy curves in the condition of $H = 0$ and 60 T. The ground state change from the paramagnetic to ferromagnetic state when $F_1(M) = F_2(M)$. The solid line shows the experimental result, and dashed spaced and dashed lines are simulated magnetizations of state~1 and~2 with $a=400$, $b=-6000$, $c=1.0 \times 10^6$, $a'=-250$, $b'=1.5\times10^4$ and $c=0$, respectively. \label{Fig:4}}
\end{figure}

Let us now introduce the Wohlfarth-Rhodes-Shimizu (WRS) theory~\cite{Wohlfarth62, Shimizu82}, which leads to IEMT with the Landau expansion of the free energy.
The free energy with the magnetic moment $M$ and the external field $H$ are written as,
\begin{equation} 
F = F_0 + \frac{1}{2}aM^2 + \frac{1}{4}bM^4 + \frac{1}{6}cM^6 -MH, \label{eq1}
\end{equation}
\begin{equation} 
H=\frac{\partial}{\partial M}F = aM + bM^3 + cM^5, \label{eq2}
\end{equation} 
where $a$, $b$ and $c$ are the expansion coefficients originated in the electron density of states and its derivatives at the Fermi level. 
The conditions of coefficients for the occurrence of the first-order IEMT and the crossover from the paramagnetic to ferromagnetic states are respectively written as 
\begin{equation} 
a>0, b<0, c>0 \ {\rm and}\ \frac{3}{16}<\frac{ac}{b^2}<\frac{9}{20} \ ({\rm IEMT}), \label{eq3}
\end{equation}
\begin{equation} 
a>0, b<0, c>0 \ {\rm and}\ \frac{ac}{b^2}\geq\frac{9}{20} \ ({\rm crossover}). \label{eq4}
\end{equation}
Yamada developed the WRS theory taking spin fluctuations into account and succeeded in explaining the maximum behavior in the temperature dependence of $\chi(T)$, which frequently observed in nearly ferromagnetic itinerant-electron metamagnet as well as present cases~\cite{Yamada93}.
Though above model based on Landau expansion is quite simple, it succeeds in leading to the single-step IEMT.
However, this theory cannot explain our results, in which the magnetization curves show double anomalies.
This model employs the free energy of a single state described by Eq.~(\ref{eq1}).
It is equivalent to the assumption that the band structure does not change in the paramagnetic and ferromagnetic states.
Removing this assumption, Takahashi and Sakai introduced a model started from two almost degenerated electronic states~\cite{Takahashi95}.
In some metamagnetic compounds such as Y(Co,~Al)$_2$ and La(Fe,~Si)$_{13}$, their band structures are changed through the metamagnetic transition and lattice parameters also change by the magnetovolume effect~\cite{Duc93, Fujita99}.
Thus using two different states is quite reasonable.
We introduce state~1 and~2, and use up to the 6th expansion term of the free energies.
The free energies of state~1, $F_1(M)$ and state~2, $F_2(M)$ are written as,
\begin{equation}
F_1(M) = F_1(0) +\frac{1}{2}aM^2+\frac{1}{4}bM^4+\frac{1}{6}cM^6-MH, \label{eq5}
\end{equation}
\begin{equation} 
F_2(M) = F_2(0) +\frac{1}{2}a'M^2 +\frac{1}{4}b'M^4+\frac{1}{6}c'M^6-MH, \label{eq6}
\end{equation}
with $a>0$, $b<0$, $c>0$, $a'<0$ and $c'\geq0$, respectively.
Figure~\ref{Fig:4}(a) shows simulated magnetization and differential magnetization curves using above equations and experimental ones of SrCo$_2$P$_2$ for comparison.
The simulated curves reproduce the small anomaly at $H_{\rm c1}$ and the first-order IEMT at $H_{\rm c2}$. 
We show a schematic free energy diagram of the present model in Fig.~\ref{Fig:4}(b).
Free energy curves of $F_1(M)$ and $F_2(M)$ show a minimum at $M_1^0$ and $M_2^0$, respectively, and the ground state magnetization is $M_1^0 = 0$ with state~1 at zero field.
Under the external field, state~2 is relatively stabilized. 
The first-order metamagnetic transition comes from the phase transformation from state~1 to~2 at $H_{\rm c2}$, where $F_1(M_1^0) = F_2(M_2^0)$ is satisfied.
The metamagnetic transition field $H_{\rm c2}$ depends on the value of $\Delta F(0)=F_2(0)-F_1(0)$.

The anomaly at lower field $H_{\rm c1}$ for $x=0$ and 0.2 occurs when the free energy parameters of state~1 in Eq.~(\ref{eq5}) satisfy the relation in Eq.~(\ref{eq4}). 
Therefore the anomaly is a kind of crossover from the ground state to somewhat higher magnetization state.
In this case, because the change of the electronic state is quite small, characteristics within state~1 model is enough to explain the experiment and one need not divide states through the crossover.

Moreover, present two-states model can explain the double maxima behavior of $\chi(T)$ and the relation between $T_{\rm max}$ and $H_{\rm c}$.
Next we discuss the temperature dependence of the free energy. 
State~1 is the ground state in the low-temperature region, and then the initially metastable state~2 is stabilized with increasing temperature and becomes the lowest energy state at $T_{\rm max}$, where $F_1(0, T_{\rm max}) = F_2(0, T_{\rm max})$ is satisfied.
Because both $H_{\rm c}$ and $T_{\rm max}$ are proportional to $\Delta F(0)^{1/2}$, $H_{\rm c}$ and $T_{\rm max}$ are predicted to show a linear relationship~\cite{Takahashi98}.
As shown in Fig.~\ref{Fig:3}(b), $H_{\rm c2}$ is almost proportional to $T_{\rm max}$ being consistent with the prediction.
Note that in the case of $x = 0$ and 0.2, there are two anomalies both in the temperature dependence of the magnetic susceptibility and the magnetization curve.
The lower-temperature maximum at $T_{\mathrm{max}1}$ is explained by the temperature dependence of expansion coefficients of $F_1(H, T)$ as well as the crossover in magnetization curves.

The simulated parameters of state~1 and 2 are written in the caption of Fig.~\ref{Fig:4}.
Takahashi and Sakai's theory requires that the two states are almost degenerated.
If our simulated result for the difference of energies between state~1 and~2 is out of theoretical requirement, it may be due to contribution of the two-band nature.
In any case, if two states have quite different magnetizations, one should treat with using separable free energies.

Next we discuss the $x$ dependence.
Similar saturation moments in the ucT region in $x\leq0.5$ suggest that the band structure of state~2 does not change markedly with $x$.
According to the above model, $\Delta F(0)$ decreases with Ca substitution as well as the metamagnetic transition field.
In this system, the lattice collapse along the $c$ axis with Ca substitution enhances the interlayer correlation of the electronic structure.
This fact is the reason why Co moments order ferromagnetically intralayer and antiferromagnetically interlayer in the cT region.

A ferromagnetic QCP is found to lies at $x=0.5$ in this system.
The first-order-like lattice collapse transition from ucT to cT phases occurs at almost the same point ($x=0.5$), and antiferromagnetic orderings take place as a ground state in cT region ($x>0.5$) by interlayer interactions through the P-P bonding. 
In the ucT region ($x<0.5$), the ferromagnetic interactions are found to increase with a positive chemical pressure by Ca substitution, which is an opposite phenomenon of the usual effect with the band narrowing.
In this system, the following two possibilities are candidates as the origin of the enhancement of ferromagnetic interactions.
The first candidate is the increase of carrier density at a mainly cobalt contributed band by the decrease of the interlayer P-P distance and the Co-P coupling, and the second one is the suppression of spin fluctuations which disturb magnetic orderings, by the interlayer interactions through the P-P coupling.

In summary, we have found novel metamagnetic transitions and ferromagnetic quantum criticality in Sr$_{1-x}$Ca$_x$Co$_2$P$_2$ ($x\leq0.5$) with the quasi-2D ucT cell.	
Especially, the compounds with $x=0$ and 0.2 show two maxima in the temperature dependence of the magnetic susceptibility.
These two anomalies have the same origin with the two anomalies in the magnetization curve, that is, the crossover and the IEMT phenomena, which cannot be explained by the conventional WRS theory.
We can succeed in quantitatively explaining our result with the extended model assuming two separated states.
The present novel results are expected to lead a breakthrough of the field of the itinerant magnetism, especially investigation of the metamagnetism in the quasi-2D system.

\begin{acknowledgments}
This work is supported by Grants-in-Aid for Scientific Research 22350029 and 23550152 from the Ministry of Education, Culture, Sports, Science and Technology of Japan and Grants for Excellent Graduate Schools, MEXT, Japan.
\end{acknowledgments}

\end{document}